\documentclass[aps,prd,amsmath,amssymb,showpacs,showkeys,nofootinbib]{revtex4}
\usepackage{graphicx}
\usepackage{dcolumn}
\usepackage{bm}
\usepackage{epsfig}

\begin{document}
\title{Accuracy of bound-state form factors extracted from
dispersive sum rules}%HEPHY-PUB 865/08
\author{Wolfgang Lucha$^{a}$, Dmitri Melikhov$^{a,b}$, and Silvano
Simula$^{c}$} 
\affiliation{ $^a$Institute for High Energy Physics,
Austrian Academy of Sciences, Nikolsdorfergasse 18, A-1050,
Vienna, Austria\\$^b$Nuclear Physics Institute, Moscow State
University, 119992, Moscow, Russia\\$^c$INFN, Sezione di Roma III,
Via della Vasca Navale 84, I-00146, Roma, Italy}
\date{\today}
\begin{abstract}
We discuss the extraction of form factors from three-point sum 
rules making use of harmonic-oscillator model, where we derive the exact 
expression for the relevant correlator. We determine the form 
factor of the ground state by the standard procedures adopted 
in the method of sum rules, and compare the obtained results with 
the known exact values. We show that the uncontrollable uncertainty in the
extracted value of the form factor is typically much larger than
that for the decay constant. In the example considered, we find
the uncontrolled systematic error in the extracted form factor to
exceed the 10\% level.
\end{abstract}
\maketitle

\section{Introduction}
A QCD sum-rule calculation of hadron parameters \cite{svz,ioffe}
involves two steps: (i) one calculates the operator product
expansion (OPE) series for a relevant correlator and formulates
the sum rule which relates this OPE to the sum over hadronic
states, and (ii) one extracts the parameters of the ground state
by some numerical procedure. Each of these steps leads to
uncertainties in the final result.

The first step lies fully within QCD and allows for a rigorous
treatment of the uncertainties: the correlator in QCD is not known
precisely (because of uncertainties in quark masses, condensates,
$\alpha_s$, radiative corrections, etc.) but the corresponding
errors in the correlator may be controlled systematically (at
least in principle).

The second step lies beyond QCD and is more cumbersome: even if
several terms of the OPE for the correlator were known precisely,
the hadronic parameters may be extracted from a sum rule only with
limited accuracy -- the corresponding error has to be treated as
a systematic error of the employed method.

In this Letter, we continue our study of the
systematic errors of hadron parameters obtained from dispersive
sum rules. In an earlier analysis \cite{lms_sr} we addressed the
determination of the decay constant of the ground state by means
of the two-point correlator. Here, we consider the extraction of
the ground-state form factor from the three-point correlator in a
quantum-mechanical harmonic-oscillator (HO) potential model. This
simple model has strong advantages compared to more complicated
cases: (i) it enables one to calculate the exact three-point
function, and thus to generate the OPE to any order, and (ii) the
bound-state parameters (masses, wave functions, form factors) are
known precisely. Therefore, we may apply the standard sum-rule
machinery to extract the form factor of the ground state and then
compare it with the known exact form factor. In this way, we may
probe the accuracy and reliability of the method. (For a
discussion of many aspects of sum rules in quantum mechanics, we
refer to \cite{nsvz,nsvz1,qmsr,orsay}.)

We present an explicit example of the form factor extracted at a
specific value of the momentum transfer, for which the exact correlator
is described by the OPE with better than 1\% accuracy and 
standard sum-rule techniques yield a form factor extremely stable
in the Borel window. However, the value obtained from the sum-rule analysis 
differs from the exact value by more than 10\%.

We therefore reinforce our previous statement that the standard
procedures adopted in the method of sum rules do not allow one to
obtain rigorous error estimates for the ground-state
characteristics. In the case of form factors extracted from
three-point correlators, the uncontrolled systematic errors may be 
considerably larger than those found in the case of decay
constants extracted from two-point sum rules.

\section{Harmonic-oscillator model}
We consider a nonrelativistic model Hamiltonian $H$ with a HO
interaction potential $V(r)$, $r\equiv|\vec r\,|$:
\begin{eqnarray}
H=H_0+V(r), \qquad H_0={\vec p}^{\,2}/2m, \qquad V(r)={m\omega^2r^2}/{2}.
\end{eqnarray}
The full Green function $G(E)\equiv(H-E)^{-1}$ and the free Green
function $G_0(E)\equiv(H_0-E)^{-1}$ are related by
\begin{eqnarray}
G^{-1}(E)-G_0^{-1}(E)=V.
\end{eqnarray}
The solution $G(E)$ of this relation may be easily found by
constructing its expansion in powers of the interaction $V$:
\begin{eqnarray}
\label{ls} G(E)=G_0(E)-G_0(E)VG_0(E)+\cdots.
\end{eqnarray}
In our HO model, all characteristics of the bound states are
easily calculable. For instance, for the ground state~($n=0$) one
finds, with $q\equiv |\vec q|,$
\begin{eqnarray}
\label{E0} E_0=\frac{3}{2}\omega,\qquad R_0\equiv |\Psi_0(\vec
r=\vec 0)|^2=\left(\frac{m\omega}{\pi}\right)^{3/2},\qquad
F_0(q)=\exp(-q^2/4m\omega),
\end{eqnarray}
where the elastic form factor of the ground state, $F_0(q),$ is
defined according to
\begin{eqnarray}
F_0(q)=\langle \Psi_0|J(\vec q)|\Psi_0\rangle=\int d^3k\,
\psi^\dagger(\vec k)\psi(\vec k-\vec q)= \int d^3r\, |\psi(\vec
r)|^2e^{i\vec q\vec r}
\end{eqnarray}
and the current operator $J(\vec q)$ is given by the kernel
\begin{eqnarray}
\label{J} \langle \vec r\,'|J(\vec q)|\vec r\rangle=\exp(i\vec
q\vec r)\delta^{(3)}(\vec r-\vec r\,').
\end{eqnarray}

\section{Polarization operator}
The polarization operator
\begin{eqnarray}
\label{pi} \Pi(T)=\langle \vec r_f=\vec 0|\exp(- H T)|\vec
r_i=\vec 0\rangle
\end{eqnarray}
is used in the sum-rule approach for the extraction of the wave
function at the origin (i.e., of the decay constant)~of~the ground
state \cite{svz}. A detailed analysis of the corresponding
procedure for the HO model was presented in \cite{lms_sr}.
For~the~HO potential, the analytic expression for $\Pi(T)$ is
known \cite{nsvz}:
\begin{eqnarray}
\label{piexact} \Pi(T)=\left(\frac{\omega
m}{\pi}\right)^{3/2}\frac1{\left[2\sinh(\omega T)\right]^{3/2}}.
\end{eqnarray}
The OPE series is the expansion of the exact quantity at small
Euclidean time $T$ (or, equivalently, in powers of $\omega$):
\begin{eqnarray}
\label{piope} \Pi_{\rm OPE}(T)=\left(\frac{m}{2\pi T}\right)^{3/2}
\left(1-\frac{1}{4}\omega^2T^2+\frac{19}{480}{\omega^4 T^4}
+\cdots \right).
\end{eqnarray}

\section{Vertex function}
The basic quantity for the extraction of the form factor in the
method of dispersive sum rules is the correlator of three currents
\cite{ioffe}. The analogue of this quantity in quantum mechanics
is
\begin{eqnarray}
\Gamma(E_2,E_1,q)= \langle \vec r_f=\vec 0|(H-E_2)^{-1}J(\vec
q)(H-E_1)^{-1}|\vec r_i=\vec 0\rangle,\qquad q\equiv |\vec q|,
\end{eqnarray}
[with the operator $J(\vec q)$ defined in (\ref{J})] and its
double Borel (Laplace) transform under $E_1\to \tau_1$ and $E_2\to
\tau_2$
\begin{eqnarray}
\Gamma(\tau_2,\tau_1,q)= \langle \vec r_f=\vec 0|\exp(-H
\tau_2)J(\vec q)\exp(-H \tau_1)|\vec r_i=\vec 0\rangle.
\end{eqnarray}
For large $\tau_1$ and $\tau_2$ the correlator is dominated by the ground state:
\begin{eqnarray}
\Gamma(\tau_2,\tau_1,q)\to |\psi_{n=0}(r=0)|^2 e^{-E_0
(\tau_1+\tau_2)}F_0(q^2)+\cdots.
\end{eqnarray}
Let us notice the Ward identity which relates the vertex function
at zero momentum to the polarization operator:
\begin{eqnarray}
\Gamma(\tau_1,\tau_2,q=0)=\Pi(\tau_1+\tau_2).
\end{eqnarray}
This expression follows directly from the relation
\begin{eqnarray}
J(\vec q=0)=1.
\end{eqnarray}
In HO model, we find the exact analytic expression for
$\Gamma(\tau_2,\tau_1,q)$ by using the results for the
Green function in configuration space $\langle \vec r\,'=\vec
0|G(T)|\vec r\rangle$ from \cite{nsvz}. For our further
investigation, we consider the vertex function for equal times
$\tau_1=\tau_2=\frac12 T,$ which has the following explicit form:
\begin{eqnarray}
\label{gamma} \Gamma(T,q)=\left(\frac{m\omega}{\pi}\right)^{3/2}
\frac{1}{\left[2 \sinh(\omega
T)\right]^{3/2}}\exp\left(-\frac{q^2}{4m\omega}
\tanh\left(\frac{\omega T}{2}\right)\right).
\end{eqnarray}
The correlator is a function of two dimensionless variables
$\omega T$ and $q^2/m\omega$.

\begin{figure}
\includegraphics[width=12cm]{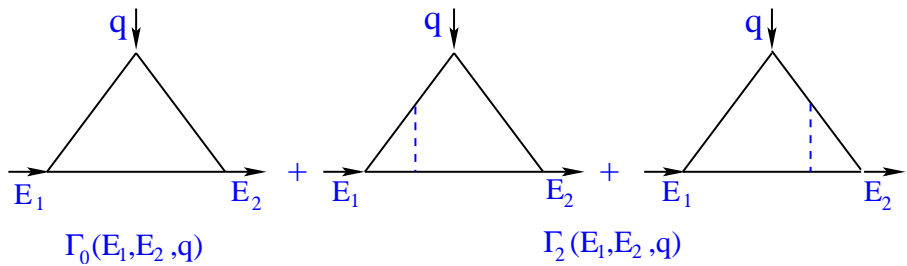}
\caption{\label{Fig:1} Expansion of the correlator
$\Gamma(E_1,E_2,q)$ in powers of the interaction.}
\end{figure}

Let us now construct for $\Gamma(T,q)$ the analogue of the OPE
as used in the method of three-point sum rules in QCD. The
corresponding procedure consists of two steps: First, we expand
$\Gamma$ in powers of $\omega^2$ and obtain
\begin{eqnarray}
\label{gamma1}
\Gamma(T,q)=\sum\limits_{n=0}\Gamma_{2n}(q,T)\omega^{2n}.
\end{eqnarray}
Each term in this expansion can be computed from the diagrams
depicted in Fig.~\ref{Fig:1}.
This is, however, not the full story: In three-point sum rules one
works with local condensates and, therefore, has for each
$\Gamma_{2n}$, $n\ge 1,$ a power-series expansion in $T$. To keep
the same track, we expand $\Gamma_{2n}$, $n\ge 1,$ in powers of
$T$.

In applications, we keep the terms $\Gamma_0,\ldots,\Gamma_6$ and
omit higher-order terms; we then obtain the power corrections by
expanding $\Gamma_2$, $\Gamma_4$, and $\Gamma_6$ in powers of $T$
retaining terms up to order $T^8$.

As the result of this procedure, the analogue of the OPE for
$\Gamma$ takes the form
\begin{eqnarray}
\label{gammaope} 
&&\Gamma_{\rm OPE}(T,q)=\Gamma_0(T,q)+\Gamma_{\rm power}(T,q), \nonumber\\ 
&&\qquad \Gamma_0(T,q)=\left(\frac{m}{2\pi T}\right)^{3/2}\exp\left(-\frac{q^2T}{8m}\right), \nonumber\\
&&\qquad \Gamma_{\rm power}(T,q)=\left(\frac{m}{2\pi T}\right)^{3/2} 
\left[ -\frac{1}{4}\omega^2T^2+\frac{q^2\omega^2}{24m}T^3 +
\left(\frac{19}{480}\omega^4-\frac{5q^4 \omega^2}{1536 m^2}\right)T^4+\cdots \right].
\end{eqnarray}
We display here only terms up to $O(T^4)$ in $\Gamma_{\rm power}$
but in calculations retain terms up to $O(T^8)$ and $O(\omega^6)$.
These terms, as well as higher-order terms, may be easily
generated from the exact expression (\ref{gamma}).

It should be emphasized that the coefficients of each power of
$T^n$ in the square brackets of (\ref{gammaope}) are
polynomials in $q^2$ of order $(n-2)$. Therefore, if the momentum
$q$ increases, one needs to include more and more power
corrections in order to have a certain accuracy of the
truncated OPE series for $\Gamma$. In QCD, this implies the
necessity to know~and include condensates of higher dimensions and
restricts the applicability of three-point sum rules to the region
of~not~too large $q^2$.

\begin{figure}
\begin{tabular}{cc}
\includegraphics[width=6.5cm]{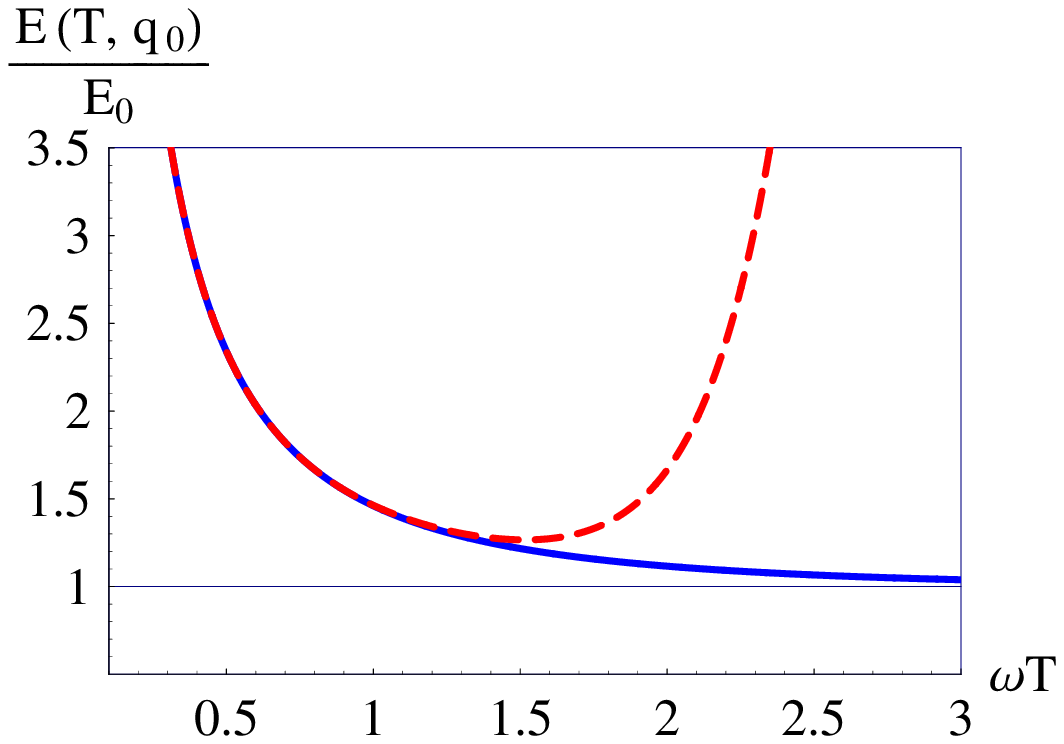}&
\includegraphics[width=6.5cm]{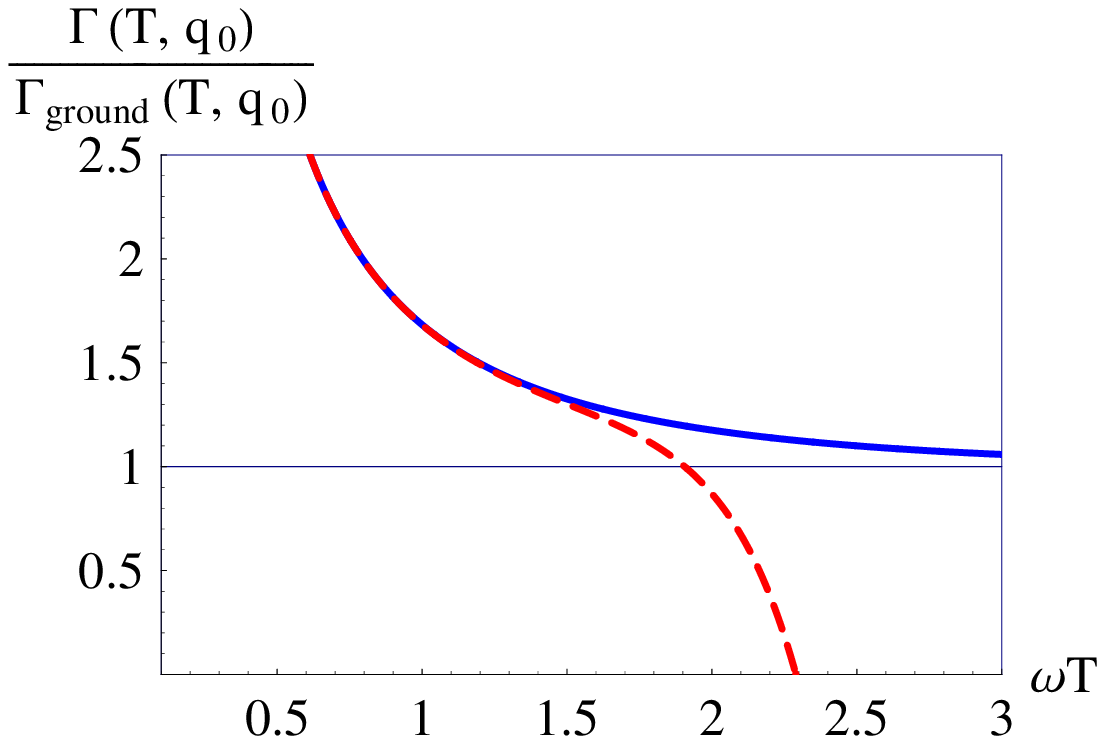}\\
(a)& (b)\\
\includegraphics[width=6.5cm]{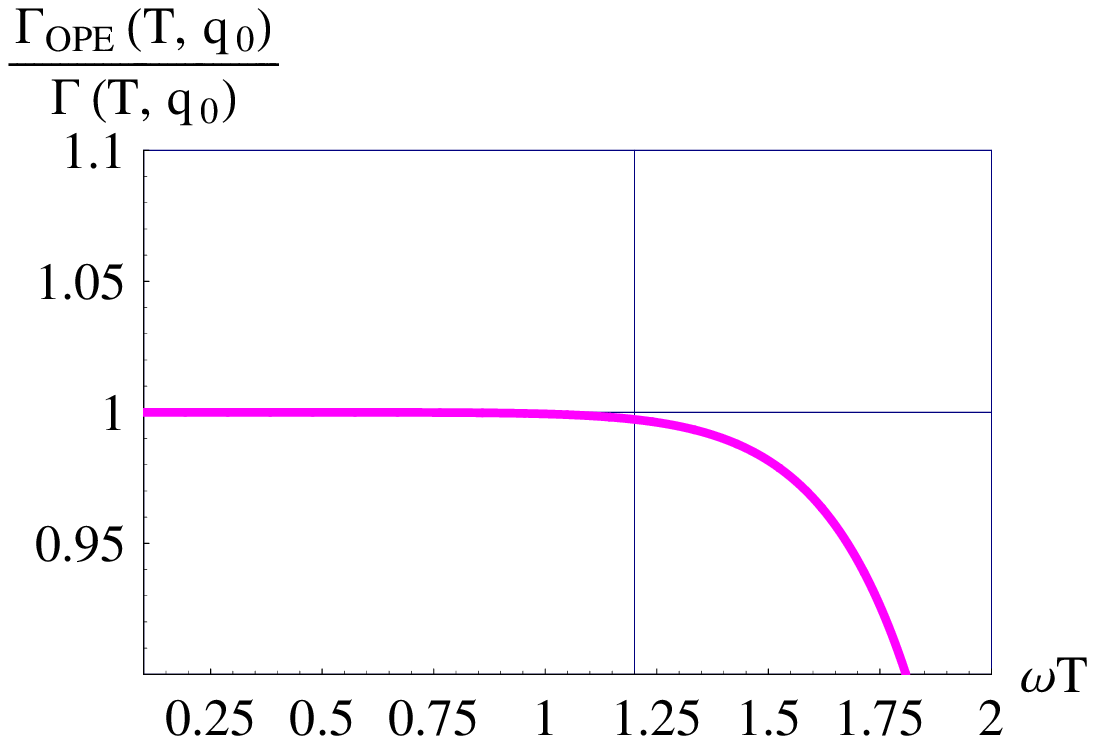}&
\includegraphics[width=6.5cm]{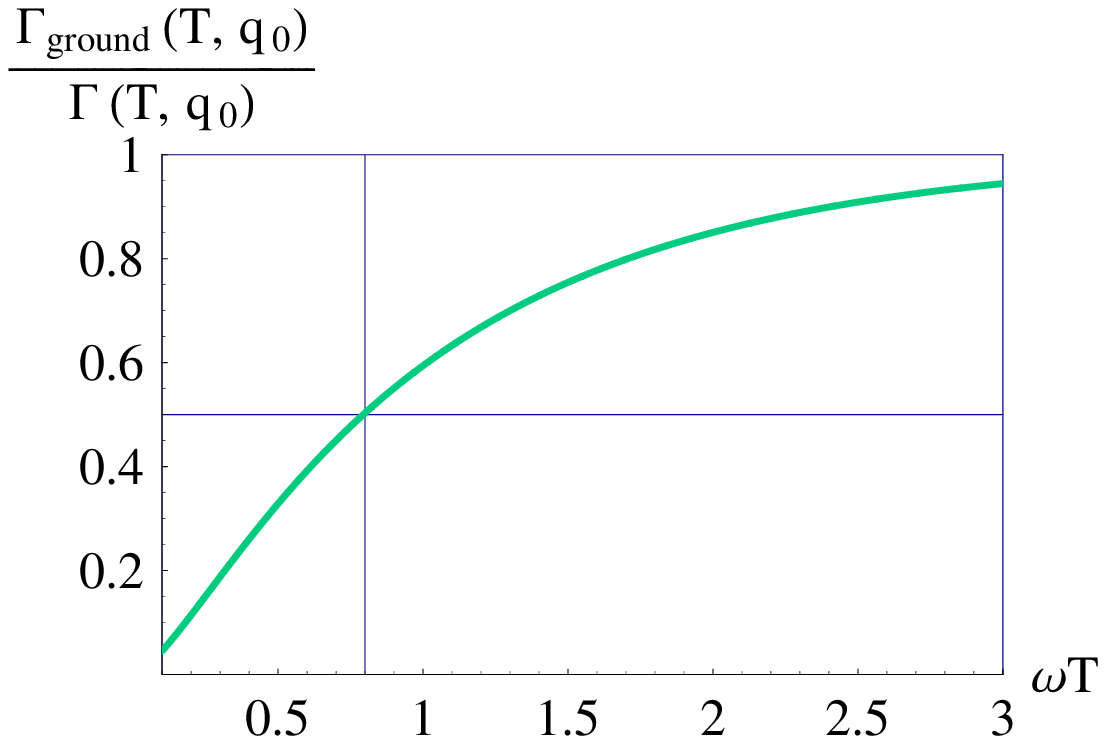}\\
(c)& (d)
\end{tabular}
\caption{\label{Plot:2.1} 
(a) The energy $E(T,q_0)=-\partial_T\log
\Gamma(T,q_0)$ and 
(b) the correlator $\Gamma(T,q_0)$ for
$q_0=1.5\,\omega$. Blue (full) line: calculation~for the exact
$\Gamma(T,q_0)$; red (dashed) line: calculation for the truncated
OPE series $\Gamma_{\rm OPE}(T,q_0)$ with corrections up to order
$(\omega T)^8$. 
(c) The accuracy of the truncated OPE for
$\Gamma(T,q_0)$ for $q_0=1.5\,\omega$. The vertical line at
$q_0=1.2\,\omega$ is the upper boundary of the Borel window. 
(d) Relative contribution of the ground state to the correlator. The
vertical line at $q_0=0.8\,\omega$ is the lower boundary of the Borel window. }
\end{figure}

Figure \ref{Plot:2.1} demonstrates the behaviour of the exact
correlator and the truncated OPE series as described above for a
fixed momentum transfer $q_0=1.5\,\omega$.
Figures \ref{Plot:2.1}a,b make obvious how the ground-state form
factor may be extracted from the correlator $\Gamma(T,q)$ known
numerically (e.g., from the lattice): The correlator is dominated
by the ground state at large values of $T$; so one may calculate
the $T$- and $q$-dependent energy
\begin{eqnarray}
\label{plateau1} E(T,q)=-\partial_T \log
\Gamma(T,q),\qquad\partial_T\equiv\frac{\partial}{\partial T},
\end{eqnarray}
which exhibits a plateau at large $T$: $E(T,q)\to E_0$ for any
$q$. Making sure that one has already reached the plateau and that
the correlator is saturated by the ground state, one obtains the
form factor from the relation
\begin{eqnarray}
\label{plateau2}
F_0(q)=\lim_{T\to\infty}\frac{1}{R_0}e^{E_0T}\Gamma(T,q).
\end{eqnarray}
Fig.~\ref{Plot:2.1}c shows that the truncated OPE of
(\ref{gammaope}) provides a good (say, better than 1\% accuracy)
description of $\Gamma(T,q)$ in the region $\omega T\le 1.2$.
The contribution of the excited states is still rather large in this region of $\omega T$ 
(see Fig.~\ref{Plot:2.1}d) and thus a direct determination of the
form factor from a truncated OPE is not possible. The procedures
of the sum-rule method are aimed at {\it modeling} the contribution of
higher states to the correlator and at obtaining in this way the
ground-state form factor.

\section{Sum rule}
The sum rule is merely an expression of equality of the correlator
calculated in the ``quark'' and in the hadron basis:
\begin{eqnarray}
R_0 e^{-{E_0}T}F_0(q)+\Gamma_{\rm excited}(T,q)=\Gamma_0 + \Gamma_{\rm power} (T,q). 
\label{sr1}
\end{eqnarray}
The quantity $\Gamma_0$ describes the free propagation and does
not depend on the interaction. It may be written as a double
spectral representation \cite{lms_prd75}:
\begin{eqnarray}
\Gamma_0(T, q)=\int dz_1 dz_2 e^{-{\frac12} z_1 T} e^{-{\frac12}
z_2 T}\Delta_0(z_1,z_2,q), \qquad
\Delta_0(z_1,z_2,q)=\frac{1}{16\pi^2 q}\theta\left((z_1+z_2-q^2)^2-4z_1z_2<0\right).
\end{eqnarray}
Making use of the standard assumption that the contribution of the
ground state is dual to the (rectangular) region~of small values
of $z_1$ and $z_2$, we obtain the relation:
\begin{eqnarray}
R_0 e^{-{E_0}T}F_0(q)
=\int\limits_{0}^{z_{\rm eff}(T,q)}dz_1 \int\limits_{0}^{z_{\rm eff}(T,q)}dz_2\,e^{-\frac12 z_1T}e^{-\frac12 z_2T}
\Delta_0(z_1,z_2,q)+ \Gamma_{\rm power}(T,q).
\label{sr}
\end{eqnarray}
The above expression is exact if we use the exact $T$- and
$q$-dependent effective continuum threshold, which cannot be
calculated from the knowledge of only the OPE but can, of course,
be reconstructed in our HO model, since we know the exact form
factor (for details, see \cite{lms_sr}). Moreover, Eq.~(\ref{sr})
may be even understood as the definition of the exact effective
continuum threshold, if one makes use of the exact hadron
parameters on the l.h.s. Therefore, this sum rule alone is {\em
not\/} predictive. The form factor (as well as any other
parameter) of the ground state may be obtained in the method of
sum rules only if one imposes an independent criterion to fix the
effective continuum threshold. It should be, however, understood
that this procedure is essentially hand-made and does not arise
from the underlying theory.

The standard assumption is to approximate $z_{\rm eff}(T,q)$ by a
$T$-independent quantity, i.e., to replace it according to $z_{\rm
eff}(T,q)\to z_c$. The quantity $z_c$ either may be chosen as a
$q$-independent constant or may be adjusted for any value~of $q$
separately.

We will now provide an example for the extraction of the form
factor from sum rules where all the standard criteria point to a
very accurate determination of the form factor; the actual error,
however, turns out to be much larger. 

Let us consider the extraction of the form factor at $q_0=1.5\,\omega$. This
specific value is chosen on purpose: for this momentum transfer
the sum rule for the form factor turns out to be most stable in
the Borel window.
\begin{figure}[h]
\begin{tabular}{cc}
\includegraphics[width=6.5cm]{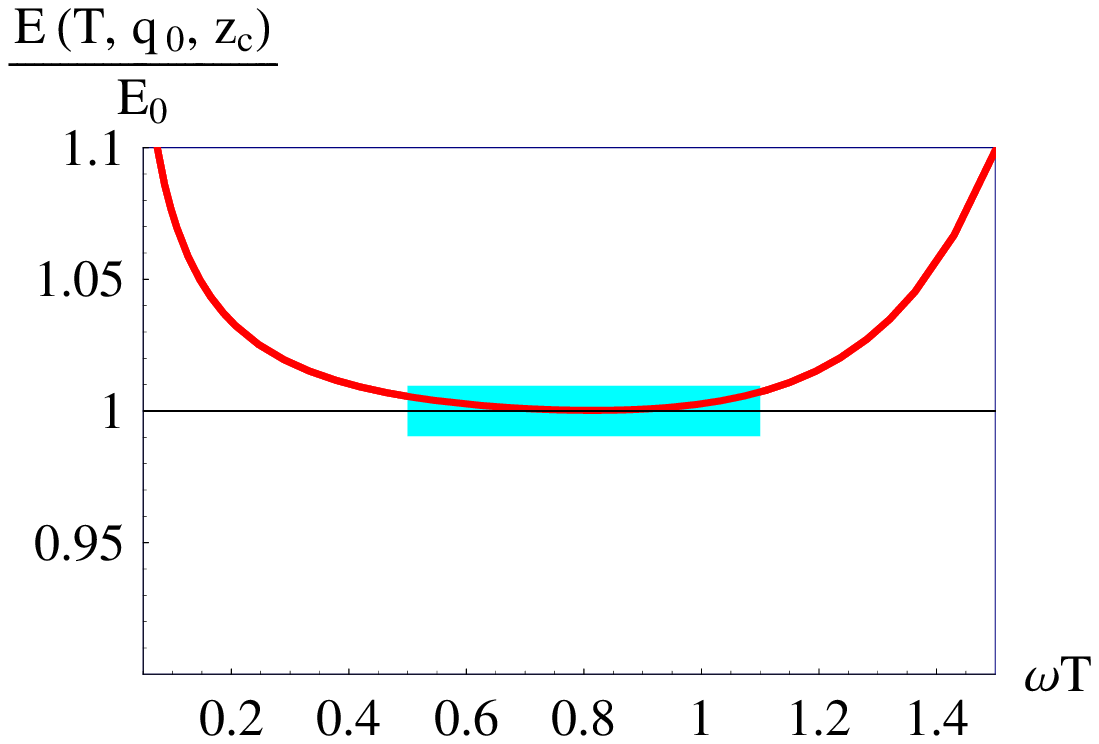}&\hspace{.2cm}
\includegraphics[width=6.5cm]{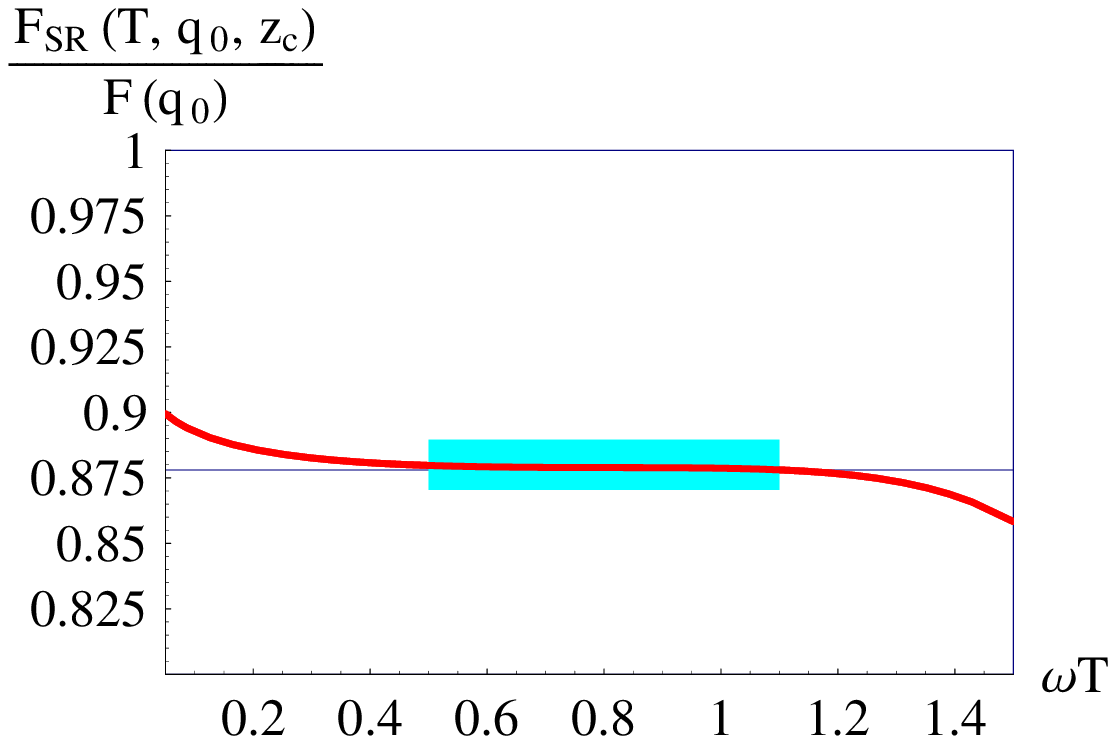}\\
(a)&\hspace{.2cm} (b)\\
\hspace{.4cm}\includegraphics[width=6.5cm]{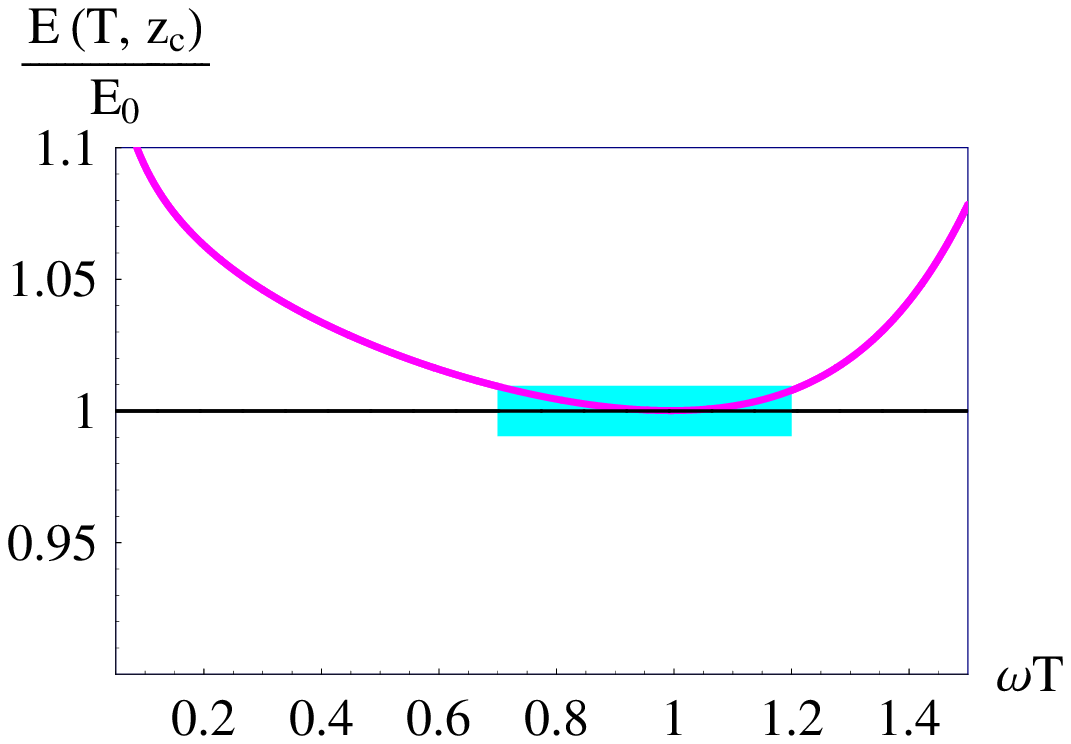}&\hspace{.2cm}
\includegraphics[width=6.5cm]{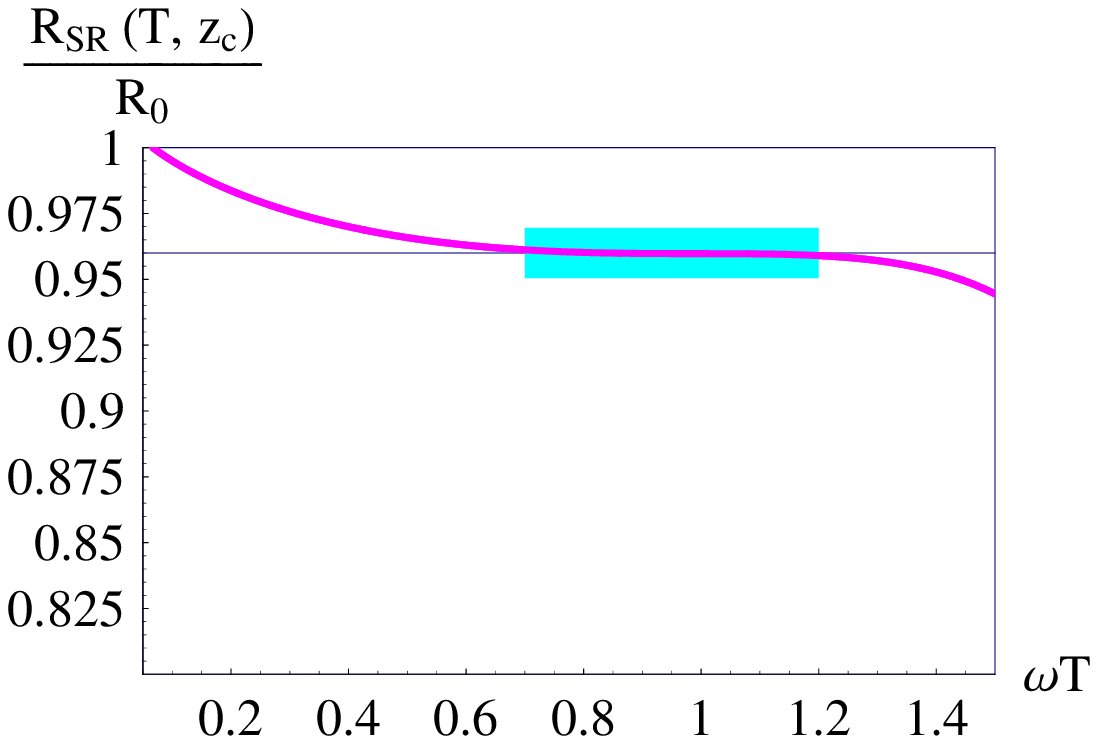}\\
(c)& \hspace{.2cm}(d)
\end{tabular}
\caption{\label{plot:2.3} 
(a) The energy of the cut correlator
$E(T,q_0,z_c)=-\partial_T\log \Gamma(T,q_0,z_c)$ and 
(b) the sum-rule estimate for the form factor at $q_0=1.5\,\omega$ for $z_c=2.42\omega$. 
(c) The energy of the two-point cut correlator $E(T,z_c)=-\partial_T\log \Pi(T,z_c)$ and 
(d) the sum-rule estimate for the parameter $R_0$ for $z_c=2.454\omega$. The shaded
rectangular areas indicate the regions of stability with 1\% accuracy.}
\end{figure}

First, let us determine the ``fiducial'' interval (or
``window'') of $T$ by the following two requirements: 
(i) The truncated OPE gives an approximation to the exact $\Gamma$
with, say, an accuracy better than 1\%. This yields $\omega T\le 1.2$. 
(ii) The ground state gives a sizeable
contribution of, say, more than 50\% to the correlator. This leads
to $0.8\le \omega T$. So the ``window'' where we will work to
extract the ground-state form factor is $0.8\le \omega T \le 1.2$.

Next, we need to impose a criterion in order to fix $z_c$. A
widely used procedure is the following \cite{jamin}: One calculates
\begin{eqnarray}
E(T,q_0,z_c)\equiv -\partial_T \log \Gamma(T,q_0,z_c),
\end{eqnarray}
which depends on $T$ because of the approximation $z_{\rm eff}(T,q_0)\to z_c$ 
(for details, consult \cite{lms_sr}). Then, one determines~$z_c$ 
such that the function $E(T,q_0,z_c)$ has a
horizontal tangent $E=E_0$, see Fig.~\ref{plot:2.3}a. 
This gives $z_c=2.42\omega$, which is used to calculate the form factor via
Eq.~(\ref{sr}) by the replacement $z_{\rm eff}(T,q_0)\to z_{c}$. 
Our results are shown in Fig.~\ref{plot:2.3}b. The form factor is perfectly flat in the 
Borel window but nevertheless turns out to be by more than 10\% lower than the known true value. 
For comparison, we present in Figs.~\ref{plot:2.3}c,d the
corresponding plots for $R_0$ extracted from the sum rule for
$\Pi(T)$ from \cite{lms_sr}. Clearly, the general picture in both
cases is similar but the deviation from the exact result is much
greater for the form factor than for the decay constant.

Let us emphasize a rather dangerous point: 
(i) a perfect description of $\Gamma(T,q_0)$ with better than 1\% accuracy in the ``window'', 
(ii) the deviation of $E(T,q_0,z_c)$ from $E_0$ at the level of only 1\%, and 
(iii) a very good stability of $F(T,q)$ with better than 1\% in the full ``window'' 
lead to an error of more than 10\% in the extracted value of $F(q_0)$! 
Clearly, this
error could not be guessed on the basis of the other numbers
obtained: the full picture mimics a very accurate extraction of
the form factor, which is, however, certainly not the case.

\newpage
\section{Conclusions}
Let us summarize the lessons one should learn from our analysis:

\noindent 1. The knowledge of the correlator in a limited
range of relatively small Euclidean times (that is, large~Borel
masses) is not sufficient for the determination of the
ground-state parameters. As a consequence, a sum-rule extraction
of the ground-state parameters {\it without knowing the
contribution of the hadronic continuum} suffers from uncontrolled
systematic uncertainties.

\noindent 2. Modeling the hadron continuum by a {\it
Borel-parameter-independent} effective continuum threshold $z_c$
allows one to fix this quantity $z_c$ by, e.g., requiring the
average energy $E(T)$ to be close to $E_0$ in the Borel window. In
this case, however, the error of the extracted ground-state
parameter turns out to be typically much larger than 

(i) the error of the description of the exact correlator by the truncated OPE and 

(ii) the variation of the bound-state parameter in the Borel window.

\noindent 3. It is important to realize that the Borel stability
of the extracted ground-state parameter --- the standard
criterion that is believed to control both the reliability and the
accuracy of the extracted ground-state parameter --- does not in fact 
guarantee the extraction of its true value.

\noindent 4. The adopted standard procedures for estimating the
errors of the extracted bound-state parameters do not allow one to
provide realistic error estimates.

The impossibility to control, at present, the systematic errors
of the extracted hadron parameters is the weak feature of the sum-rule method and an
obstacle for using the results from QCD sum rules for precision
physics, such as electroweak physics.

Finally, we would like to comment on the obtained quantitative
estimates. In HO model, the ground state is well separated
from the first excitation, which contributes to the correlator, by
a large gap of $2\omega$. This makes the HO model a very
favourable case for the application of sum rules. Whether or not a
comparable accuracy may be achieved in QCD, where this feature is
absent, is questionable.

\vspace{.5cm}

\noindent {\it Acknowledgments:} The authors are grateful to Hagop
Sazdjian for interesting discussions. D.~M. would like to thank
the theory group of the Institut de Physique Nucl\'eaire,
Universit\'e Paris-Sud for hospitality during his stay in Orsay.
D.~M. gratefully acknowledges financial support from the Austrian
Science Fund (FWF) under project P17692, RFBR under project
07-02-00551, and CNRS.

\end{document}